\documentclass[aps,prl,twocolumn,longbibliography,superscriptaddress]{revtex4-1}
\usepackage{graphicx} 
\usepackage{epstopdf}
\usepackage{amsmath}
\usepackage{color}
\usepackage{multirow}
\usepackage{amssymb}
\usepackage{dcolumn}%
\usepackage{bm}%
\usepackage{ctable}
\usepackage{tabularx}
\usepackage{graphicx}

\begin{document}

\begin{titlepage}

\title[]{\Large \bf Dependence of the glass transition and jamming densities on spatial dimension}

\author{Monoj Adhikari}
\affiliation{Theoretical Science Unit, Jawaharlal Nehru Centre for Advanced Scientific Research, Jakkur Campus, 560064 Bengaluru, India}
\author{Smarajit Karmakar}
\affiliation{TIFR Center for Interdisciplinary Science, Tata Institute of Fundamental Research, RR District, Hyderabad, 500046, Telangana, India,}
\author{Srikanth Sastry}
\email[Corresponding author: ]{sastry@jncasr.ac.in}
\affiliation{Theoretical Science Unit, Jawaharlal Nehru Centre for Advanced Scientific Research, Jakkur Campus, 560064 Bengaluru, India}

\begin{abstract}
We investigate the dynamics of soft sphere liquids through computer simulations for  spatial dimensions from $d =3$ to $8$, over a wide range of temperatures and densities. Employing a scaling of density-temperature dependent relaxation times, we precisely identify the density $\phi_0$ which marks the ideal glass transition in the hard sphere limit, and a crossover from sub- to super-Arrhenius temperature dependence. The difference between $\phi_0$ and the athermal jamming density $\phi_J$,  small in 3 and 4 dimensions, increases with dimension, with $\phi_0 > \phi_J$ for $d > 4$. We compare our results with recent theoretical calculations. 
\end{abstract}

\maketitle

\end{titlepage}

\noindent{\it Introduction:} Fluid states of matter can transform to rigid, amorphous solids through the glass transition or the jamming transition. The glass transition describes the transition to disordered solid states typically in molecular systems upon a decrease in temperature, whose nature has been intensely investigated over several decades \cite{debenedetti2001supercooled,binder2011glassy,berthier2011theoretical,karmakar2014growing,royall2018race}. The jamming transition has likewise been widely investigated in {\it athermal} systems, typified by granular matter \cite{liu2010jamming,RevModPhys.82.3197,behringer2018physics}. Their relationship has also been the subject of considerable research \cite{charbonneau2017glass}. Whereas molecular glass formers and granular matter represent cases which exhibit one or the other of these phenomena, several systems, such as colloidal suspensions, in principle exhibit both phenomena, and their interplay is important, {\it e.g.} in their rheolgoy \cite{ikeda2012unified}. An idealised system in which both phenomena have been investigated in detail is the hard sphere system. Theoretical investigations over the last decade, extending the framework of the random first order transition (RFOT) theory \cite{kirkpatrick1987dynamics,kirkpatrick1988comparison,kirkpatrick1989scaling}, have focused on the hard sphere system, and a unified mean field description of both these phenomena have been developed in the limit of infinite dimensions \cite{charbonneau2017glass,parisi_urbani_zamponi_2020,mangeat2016quantitative}. These developments have naturally led to investigations of how the infinite dimensional results relate to behaviour in finite dimensions. An appealing and systematic approach to addressing questions in this regard is to study the effect of spatial dimensionality on the glass transition and jamming phenomenology, which have been pursued for hard particle systems extensively  \cite{charbonneau2011glass,charbonneau2012dimensional,charbonneau2013dimensional,charbonneau2017glass,mangeat2016quantitative,Joyjit2020,charbonneau2021memory,PC2022}. In particular, the relationship between the glass transition and jamming transition has been investigated \cite{charbonneau2011glass,mangeat2016quantitative}. In addition to systems of hard spheres, a small number of other studies have investigated the role of dimensionality in determining aspects of glassy dynamics \cite{eaves2009spatial,sengupta2012adam,sengupta2013breakdown,xu2016entropy,adhikari2021jpcb}, such as dynamical heterogeneity in a  binary mixture of the Lennard-Jones particles as a function of temperature. A more extensive investigation of the dependence on spatial dimensionality in systems where both thermal and density effects play a role are thus of great interest. In the present work, we study soft sphere assemblies interacting with a harmonic potential by investigating the dynamics at different densities and temperatures.

In the zero-temperature limit, the behavior of this system approaches the density-controlled hard sphere model, while it is similar to thermally driven fluids at high density and finite temperature. 
In the hard sphere limit, the system \emph{jams}, losing the ability to flow, at a critical density, $\phi_J$, {\it via} the non-equilibrium jamming transition \cite{Ohern2003PRE,chaudhuri2010jamming}. Several works \cite{speedy1998random,speedy1996distribution,krzakala2007landscape,mari2009jamming,chaudhuri2010jamming,Hermes2010JammingOP,Ciamarra2010DisorderedJP,ozawa2012jamming} have considered and demonstrated the scenario that the jamming density is not unique, but can occur over a range of densities. In turn, the range of jamming densities is associated with a line of glass transition densities (kinetically determined or otherwise, as in mean field results \cite{charbonneau2017glass,parisi_urbani_zamponi_2020}) ending with a {\it Kauzmann} density $\phi_0$, which may be expected to be the relevant density for the divergence of relaxation times. The relationship between the jamming and Kauzmann densities have been investigated, with varying conclusions regarding the relative values of $\phi_J$ and $\phi_0$ \cite{berthier2009compressing,berthier2009glass,charbonneau2011glass,coniglio2017universal,berthier2016equilibrium}. Several studies \cite{mari2009jamming, ciamarra2010recent, kurchan2013exact, coniglio2017universal,charbonneau2017glass,mangeat2016quantitative} also indicate that the relationship between these two transition densities depends on dimensionality. 



\begin{figure*}[htp]
\centering
\includegraphics[width=0.99\textwidth]{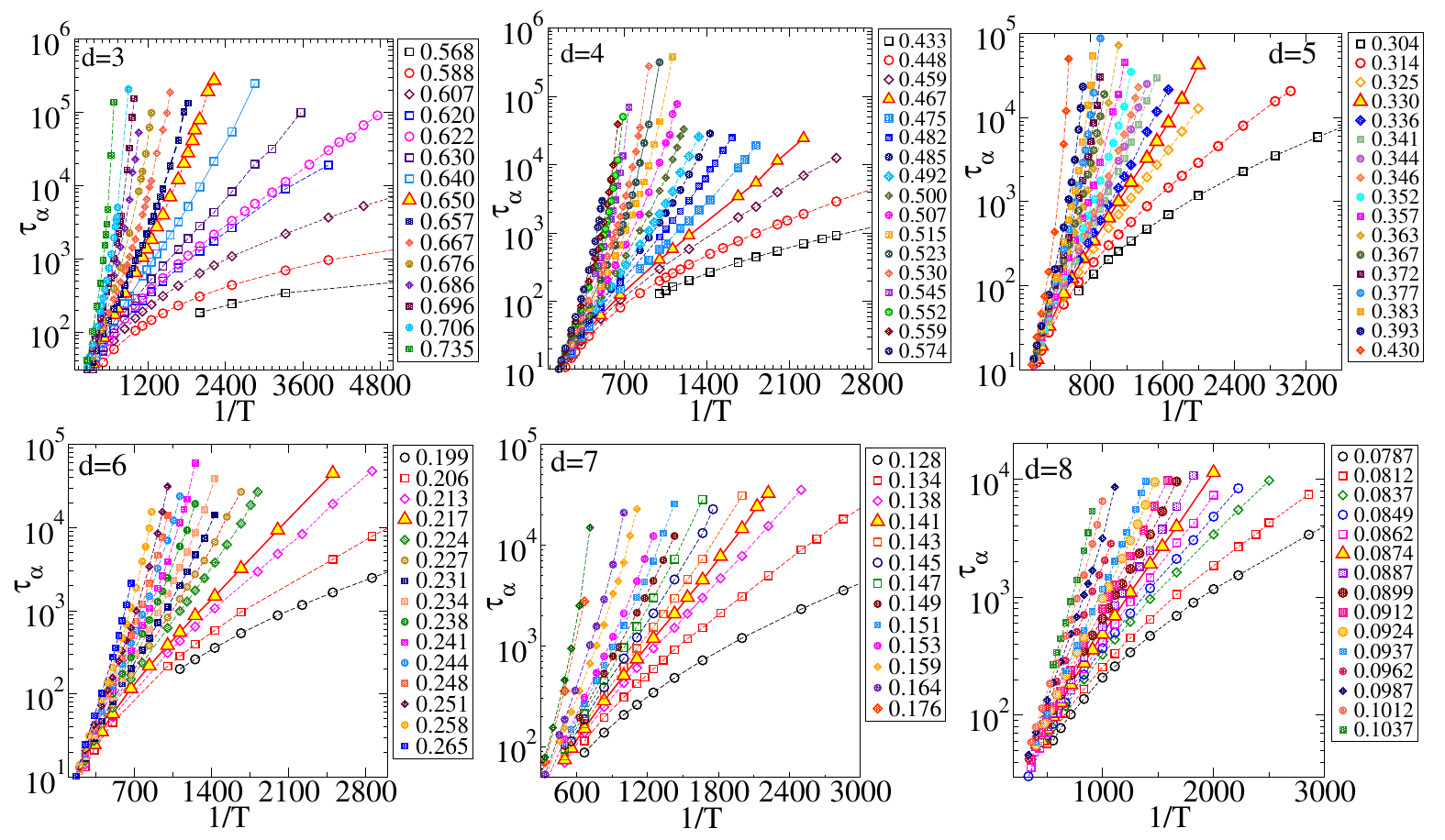}
\vskip -0.1in
\caption{Relaxation times as a function of inverse temperature are plotted in a semi-log scale for several densities in different dimensions. Data shown with  filled symbols correspond to densities at which super-Arrhenius dependence is observed.}
\label{fig.s3}
\end{figure*}

In \cite{berthier2009compressing,berthier2009glass}, the relaxation times were studied for the same model we consider, in three dimensions. With increasing density, relaxation times exhibit a cross over from sub-Arrhenius to super-Arrhenius temperature dependence. Relaxation times were analysed through a scaling function that assumes a divergence for the hard sphere systems at a density $\phi_0$, and by defining an effective hard sphere diameter at finite temperatures \cite{cohen1959molecular}, to obtain two distinct scaling collapses across $\phi_0$. The estimate of $\phi_0$ thus obtained is found to be very close to $\phi_J$, the jamming point, although the meaning of the two densities was clearly distinguished. 


In the present work, we perform extensive molecular dynamics simulations for a wide range of  $\phi, T$ values in different spatial dimensions ranging from $d = 3$ to $8$. We perform a scaling analysis similar to \cite{berthier2009compressing,berthier2009glass} but with a newly proposed scaling function, to obtain $\phi_0$ as a function of $d$. We obtain jamming densities following the analysis in \cite{Ohern2003PRE,chaudhuri2010jamming}. Our results clearly demonstrate that $\phi_0 > \phi_J$ for $d > 4$, with $\phi_0/\phi_J$ increasing with $d$, as may be expected from mean field results \cite{charbonneau2017glass,parisi_urbani_zamponi_2020,mangeat2016quantitative}.\\



\noindent{\it Simulations details:}
We study a $50:50$ binary mixture of spheres of size ratio $1.4$, interacting with a harmonic potential as a model glass former \cite{durian1995foam}:
\begin{eqnarray}
V_{\alpha \beta}(r) &=& \frac{\epsilon_{\alpha \beta}}{2}\left(1-\frac{r}{\sigma_{\alpha \beta}}\right)^{2}, \hspace{1.58cm} r_{\alpha \beta} \leq \sigma_{\alpha \beta}
\end{eqnarray}
and $V_{\alpha \beta}(r) =0$ for $r_{\alpha \beta } > \sigma_{\alpha \beta}$, where $\alpha, \;  \beta$ $\in$ (A,B), indicates the type of particle. The system size varies from $1000$ to $5000$ depending on the spatial dimensions, with the linear extent of the simulated volume, $L > 2 \sigma_{BB}$ in all dimensions. 
We investigate the dynamics at $10-14$ densities (with 1 - 2 independent samples each) around the jamming density. The volume fraction, or density,  $\phi= \rho V_d$ where $V_d =  2^{-d}\frac{\pi^{d/2}}{\Gamma(1+\frac{d}{2})}((c_A\sigma^d_{AA} + c_B\sigma^d_{BB})$, $c_A = c_B = 1/2$, is the average volume per sphere in $d$ dimensions, $\rho = {N \over V}$ is the number density, $N$ is the number of particles, and $V$ is the volume. Molecular dynamics (MD) simulations are performed in a cubic box with the periodic boundary conditions, employing the constant temperature integration in \cite{brown1984comparison}, with time step $dt = 0.01$. Each independent run is of length $ > 100\tau_\alpha$ where relaxation time $\tau_\alpha$ is computed from the overlap function $q(t)$ (for $B$ particles) as $\langle q(t = \tau_\alpha)\rangle = 1/e$, with 
\begin{equation}
q(t) = {1 \over N_B} \sum_{i=1}^{N_B} w(|\textbf{r}_j(t_0)- \textbf{r}_i(t+t_0)|),  
\end{equation} 
where $w(x) = 1$ if $x < a$ and $0$ otherwise.  In Fig. \ref{fig.s3}, the relaxation time is plotted as a function of temperature for various densities for $3-8$ dimensions. Additional details are provided in the Supplemental Material (SM) \footnote{see Supplemental Material at {\tt <URL>} which contains a description of the procedure used to define the overlap function, the scaling analysis using the scaling function defined by Berthier and Witten, the new effective diameter definition we employ in this work and the corresponding scaling function, comparison of results from the Berthier-Witten scaling function and the one proposed in this work, error analysis for different data collapse procedures employed, details of the procedure used for obtaining jamming densities, comparison of jamming densities obtain with previous work, the density-temperature phase diagram for all the dimensions investigated, and the statistics of rattlers, which contains the additional reference \cite{lavcevic2003spatially}. }


\begin{figure*}[htp]
\vskip -0.1in
\includegraphics[width=0.995\textwidth]{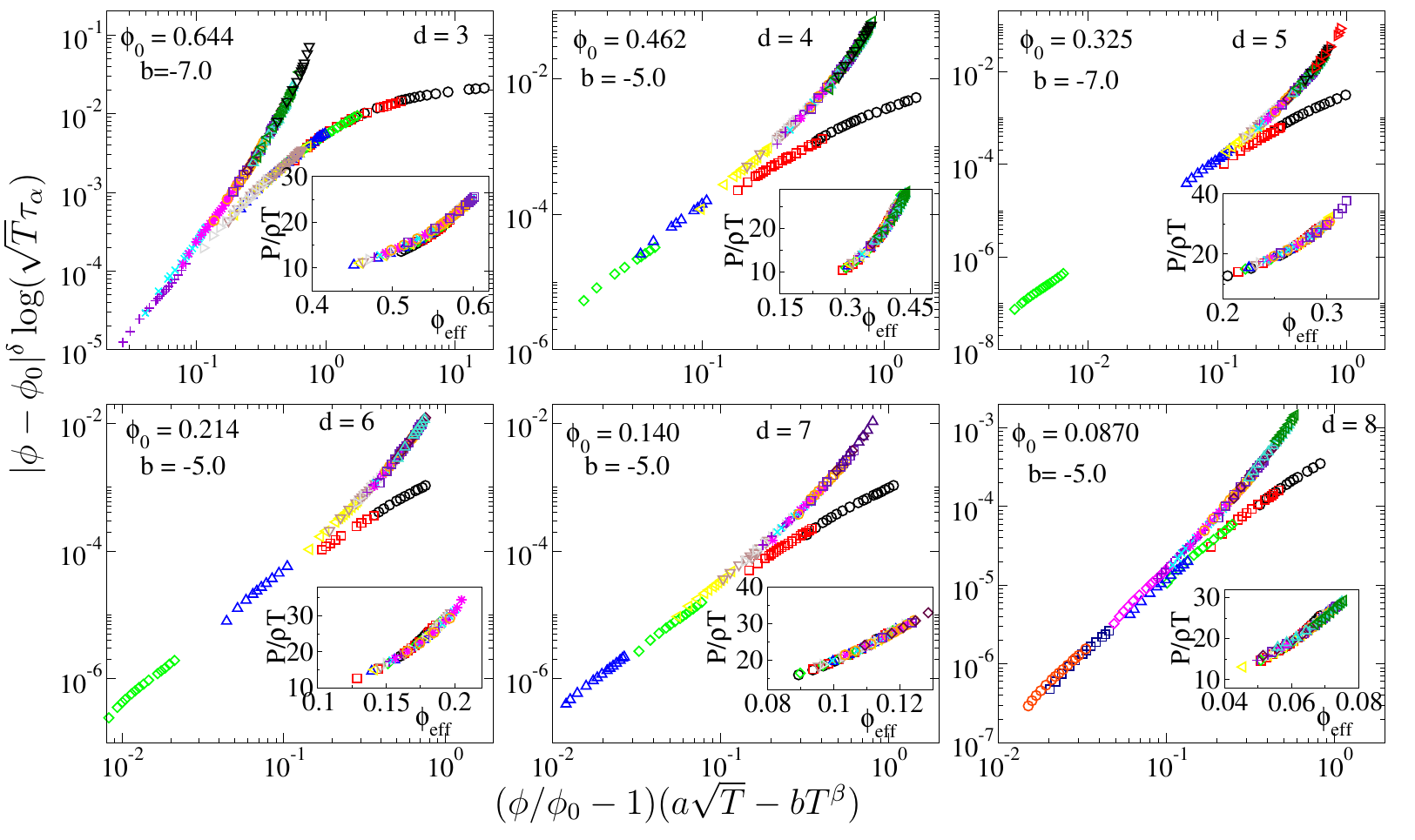}
\vskip -0.1in
\caption{Scaling plot of relaxation times for $3-8$ spatial dimension. As explained in the text, coefficients $a,b$ and exponent $\beta$ describe the effective volume fraction $\phi_{eff}$. Of these, we use fixed values of $a = d\sqrt{\pi}/2$ (see SM) and vary $b$, $\beta>0.5$ to obtain collapse of pressures on to a single curve, $P/\rho T = f(\phi_{eff})$ (shown in the insets). Best data collapse is obtained around $\beta=0.7$ (which we keep fixed), and the $b$ values are indicated in the legends. Keeping  $\delta = 2$ fixed, we vary $\phi_0$ as the single fit parameter to obtain scaling plots of relaxation times $\tau_{\alpha}$ shown. }
\label{figColl}
\end{figure*}

\noindent{\it Results:}
Considering an expression for relaxation times for hard sphere fluids of the form  $\sqrt{T}\tau_\alpha^{hs} \sim \exp \left[\frac{A}{(\phi_0 - \phi)^{\delta}}\right]$, Berthier and Witten ~\cite{berthier2009compressing} analysed relaxation times for soft spheres by defining a temperature dependent effective volume fraction of the form $\phi_{eff} = \phi - a T^{\mu/2}$, which leads to the scaling form 
\begin{equation} \label {eqnBWScaling}
\sqrt{T}\tau_{\alpha}(\phi, T) \sim \exp \left[\left (\frac{A}{|\phi_0-\phi|^{\delta}}\right)F_{\pm}\left (\frac{|\phi_0-\phi|^{\frac{2}{\mu}}}{T}\right) \right],
\end{equation}
where $F_{\pm}(x)$ refers to the scaling function for the super-Arrhenius and the sub-Arrhenius branch of the scaling function and $\mu$, $\delta$ and
$\phi_0$ are adjustable parameters. Plotting  $|\phi_0-\phi|^{\delta} \log(\tau_{\alpha})$ against $|\phi_0-\phi|^{\frac{2}{\mu}}/T$, with suitable choices of the parameters, a data collapse on to two branches above and below $\phi_0$ is obtained. The values of $\phi_0 = 0.635$ and $\delta = 2.2$ were determined from such  a procedure, with the $\delta$ value being in close agreement with experimental and simulation results for colloidal hard spheres and theoretical results ~\cite{hall1987aperiodic,brambilla2009probing,baranau2020relaxation}. The estimated $\phi_0$ is close to but distinct from the jamming density of $\phi_J = 0.648$ estimated for the binary mixture studied in \cite{Ohern2003PRE,chaudhuri2010jamming,berthier2009compressing} and here. 


Recently,  the same scaling analysis has been revisited in \cite{maiti2019temperature}, with the conclusion that if one considers $\phi_0$, $\delta$ and $\mu$ as free parameters, a good scaling collapse of comparable quality to \cite{berthier2009compressing} could be obtained for significantly different sets of parameter values. Further, the scaling form in Eq. \ref{eqnBWScaling} requires  $\mu \delta/2 =1$ in order to obtain the expected Arrhenius form at density $\phi_0$, but imposing such a constraint does not lead to the best data collapse \cite{berthier2009compressing}, as we independently verify. It is thus desirable to explore alternate scaling functions, which we do in this work based on the evaluation of an effective diameter following the prescription in \cite{barker1967perturbation}. 
We compare our results to analysis employing Eq. \ref{eqnBWScaling}, and obtain consistent estimates of $\phi_0$.

Following \cite{barker1967perturbation}, the expression for the effective diameter with only temperature dependent corrections can be written as  $\sigma_{eff}= \int_0^\sigma \left[1-\exp(-\beta u(r))\right]dr$, leading to $\sigma_{eff}  \approx  \sigma \left[ 1- \frac{1}{2}\sqrt{\pi/\beta} \right]$. In turn, the effective volume fraction in dimension $d$ can be approximated as 
\begin{equation} \label{effv_new} 
\phi_{eff}  \approx  \phi \left(1-a~\sqrt T +  b ~T^{\beta} \right)
\end{equation}
where $a = d\sqrt{\pi}/2$, and the term $bT^{\beta}$ approximates terms of ${\cal O} (T) $ and higher. Employing the effective $\phi$ in the expression for the hard sphere relaxation times, we write 
\begin{equation} \label{new-scaling}
\sqrt{T}\tau_{\alpha}(\phi, T) \sim \exp \left[ \frac{A}{|\phi_0-\phi|^{\delta}}F_{\pm}\left (\frac{|\frac{\phi_0}{\phi}-1|}{a\sqrt{T}-bT^{\beta}}\right)\right].
\end{equation}
An Arrhenius form of the relaxation times at finite $T$ and $\phi = \phi_0$ requires  $\delta$ to be $2$, which we employ. Rather than estimate $b$, $\beta$ and $\phi_0$ through a scaling analysis of $\tau_{\alpha}$, we first require that $P/\rho T$ where $P$ is the pressure is a unique function of $\phi_{eff}$ and estimate $b$ and $\beta$ from the data collapse of $P/\rho T$ (shown in the insets of Fig. \ref{figColl}). Scaling collapse of $\tau_{\alpha}$ through Eq. \ref{new-scaling}, shown in Fig. \ref{figColl}, is used to estimate the remaining parameter $\phi_0$. Additional details regarding the data collapse procedure including error analysis and justification of the choice $\delta = 2$, are provided in the SM. 

\begin{figure}[htp]
\centering
\includegraphics[width=0.49\textwidth]{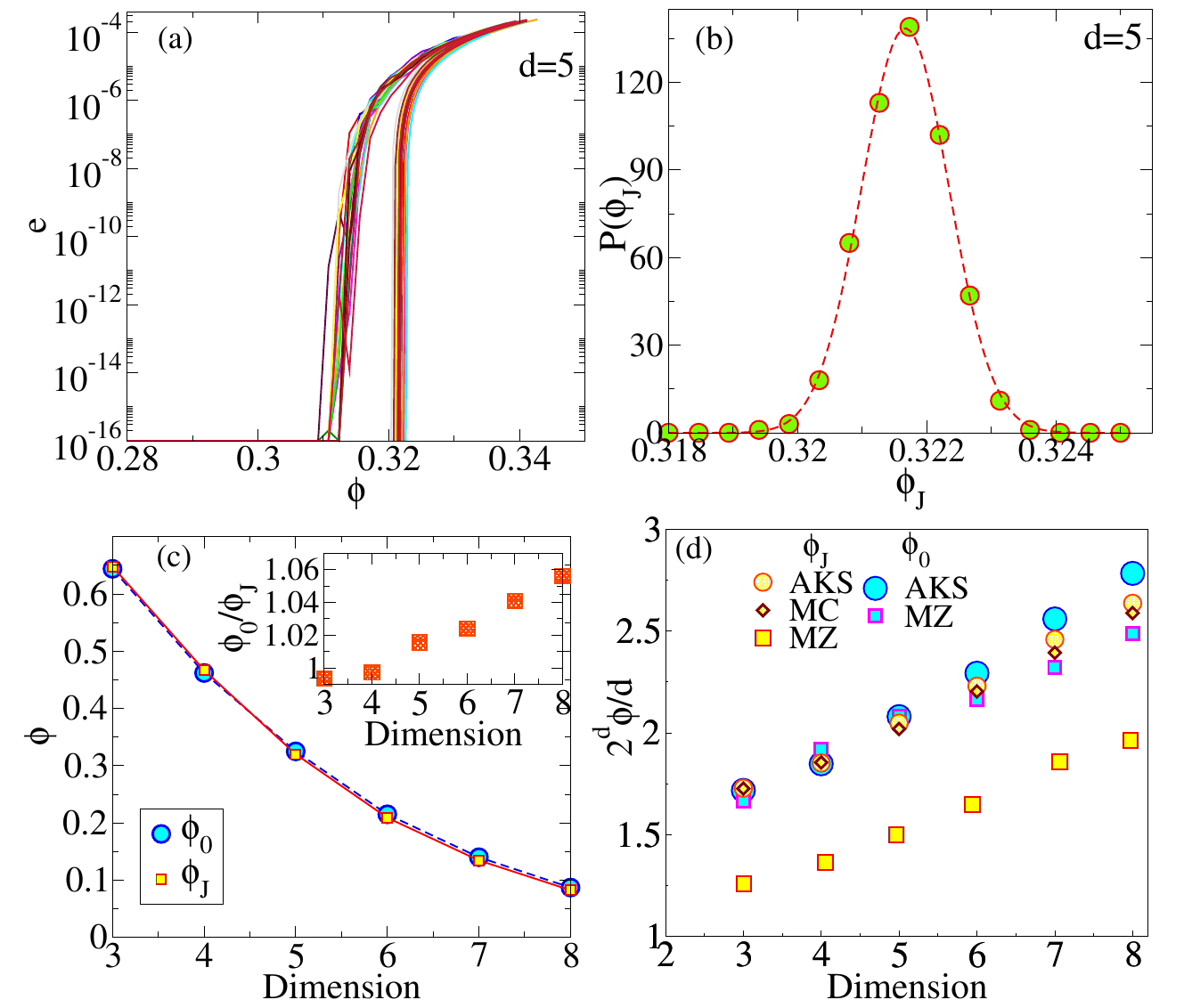}
\caption{(a): The energies during compression-decompression cycles of energy-minimized configurations plotted as a function of volume fraction for $d = 5$. Data are shown for $50$ independent samples. (b) The histogram of jamming densities for $d = 5$. (c) The jamming density, $\phi_J$ and the glass transition density, $\phi_0$, plotted as a function of spatial dimension. Inset: Ratio of $\phi_0$ and $\phi_J$,  plotted as a function of spatial dimension. (d) Comparison of $\phi_0$ and $\phi_J$ values from the present work (AKS) with previous simulation \cite{Morse2014} (MC) and theoretical calculations \cite{mangeat2016quantitative} (MZ) using the Percus-Yevick closure.}
\label{fig-jamming-5d}
\end{figure}

We next estimate the jamming densities following the protocol employed in 
\cite{chaudhuri2010jamming}, wherein initially random configurations are compressed till the energy reaches $10^{-5}$, and decompressed till the energy decreases below $10^{-16}$, at the jamming density (see SM for additional details). Fig. \ref{fig-jamming-5d}(a) illustrates the protocol  for $d = 5$ (corresponding data for all dimensions are shown in the SM). Based on the initial configuration used, the $\phi_J$ we estimate corresponds to the low density limit of the range over which jamming can take place \cite{Ohern2003PRE,chaudhuri2010jamming}.
The procedure is applied to $1000$ independent initial configurations, and the histogram of jamming densities obtained is shown in Fig. \ref{fig-jamming-5d}(b).
The average jamming density $\phi_J$ is shown in Fig. \ref{fig-jamming-5d}(c) (also tabulated in Table I), along with estimates of $\phi_0$ obtained above. The ratio $\phi_0/\phi_J$, shown in the inset, increases with $d$, with  $\phi_0 > \phi_J$ for $d > 4$, whereas for $d = 3,4$, $\phi_0 < \phi_J$, with the two values being very close. The jamming densities $\phi_J$ we obtain are very similar to, but slightly larger than, those obtained for monodisperse spheres in \cite{Morse2014,charbonneau2021memory}. In Fig. \ref{fig-jamming-5d}(d), we compare the scaled jamming densities we obtain with those in \cite{Morse2014}. We also show the recent theoretical calculations in \cite{mangeat2016quantitative} for the corresponding quantity $\phi_{th}$, which shows the same trend as our $\phi_J$ data, but are smaller (reasons why this may be expected are discussed in \cite{manacorda2022gradient}). We further show the $\phi_0$ values we obtain, along with the corresponding calculated values (Kauzmann density $\phi_K$) in \cite{mangeat2016quantitative}. As discussed in \cite{parisi_urbani_zamponi_2020,mangeat2016quantitative}, $2^d \phi_0/d$ is expected to increase as $\log d$, whereas $2^d \phi_J/d \rightarrow 6.2581$ as $d \rightarrow \infty$. We are not able to quantitatively comment on either prediction, but we note that the values of $\phi_0$ calculated in \cite{mangeat2016quantitative} are in near quantitative agreement with our results, whereas the $\phi_J$ values in \cite{mangeat2016quantitative} underestimate our results as well as those in \cite{Morse2014,charbonneau2021memory}. 



\begin{table}[h!]
\centering
\begin{tabular}{|c|c|c|}
\hline
	d &$\phi_J$ & $\phi_0$  \\
\hline
	3 & $0.648 \pm 0.0014$ & $0.644 \pm 0.003$  \\
\hline 
	4 & $0.467 \pm 0.0015$ & $0.462 \pm 0.004$ \\ 
\hline
	5 &$0.320 \pm 0.0008$ & $0.325 \pm 0.003$   \\ 
\hline
	6 & $0.209 \pm 0.0006$ & $0.214 \pm 0.002$  \\
\hline
	7 & $0.1345 \pm 0.0004$ & $0.140 \pm 0.002$ \\
\hline
	8 & $0.0824 \pm 0.0003$ & $0.0870 \pm 0.001$  \\ 
\hline
\end{tabular} 
\caption{Jamming and glass transition densities $\phi_J$ and $\phi_0$ for dimensions $d$ from $3$ to $8$. Error bars for $\phi_0$ are obtained by considering an increase in the error $\chi_{\tau}^2$ (defined in SM) by $20\%$ of the lowest value. The error bars of $\phi_J$ is computed from half width at half maximum of the distribution of $\phi_J$ (shown Fig. 3(b) and the SM).}.
\label{table-phij}
\end{table}
\begin{figure}[htp]
\centering
\includegraphics[width=0.47\textwidth]{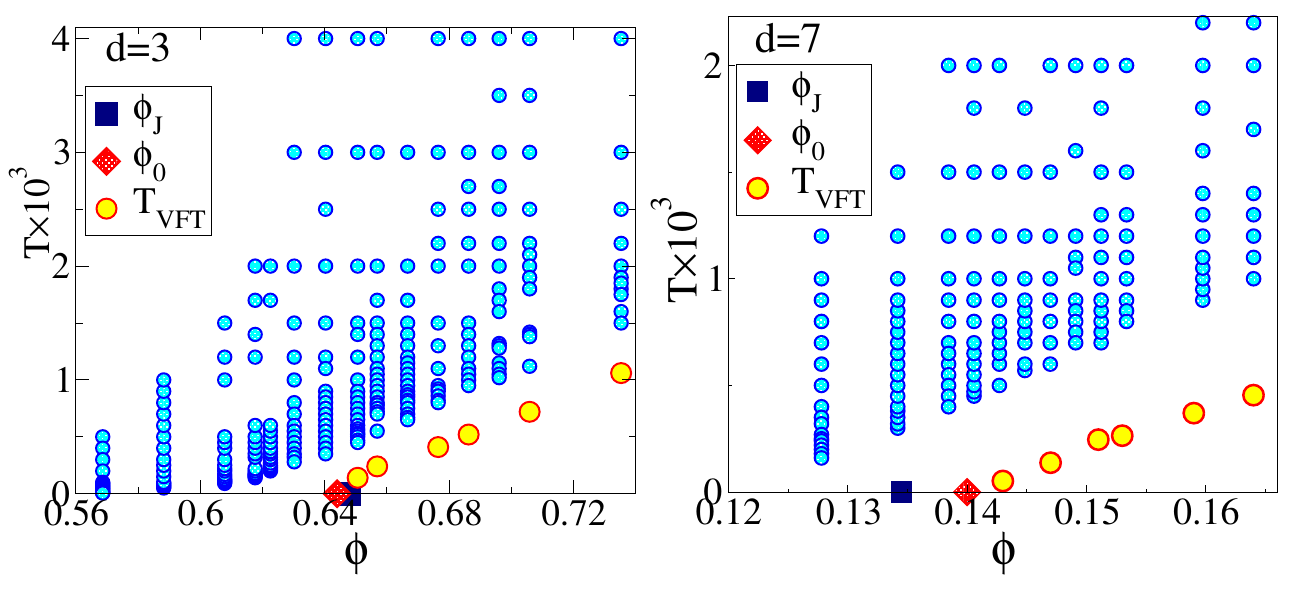}
\caption{The jamming and glass transition densities $\phi_J$ (black square) and $\phi_0$ (red diamond) are shown for $d = 3$ and $d = 5$ along with the density dependent glass transition temperatures $T_{VFT}$ (orange circles), and the set of densities and temperatures at which simulations have been performed (blue circles).}
\label{fig.s5}
\end{figure}

Finally, we compute the temperature at which the relaxation times show an apparent divergence by fitting the data at each density above $\phi_0$, for each dimension, to the Vogel-Fulcher-Tammann (VFT) form, $\tau_{\alpha} = \tau_0 \exp\left[\frac{1}{K_{VFT}\left(\frac{T}{T_{VFT}}-1\right)}\right]$.
In Fig. \ref{fig.s5}, we show the density-temperature diagram for $d = 3$ and $d = 7$ (results for other dimensions are shown in the SM) which shows $\phi_0$ and $\phi_J$, along with the density dependent $T_{VFT}$. The results shown illustrate the manner in which the relationship between $\phi_0$ and $\phi_J$ changes with spatial dimension.  The $T_{VFT}$ values shown extrapolate to zero at $\phi \rightarrow \phi_0$, illustrating  that   $\phi_0$ is the relevant limit density for the density dependent glass transition. 
 

\noindent{\it Conclusion:}
To summarize, we have studied the dynamics of model glass forming liquids consisting of soft (harmonic) spheres by measuring relaxation times as a function of temperature for several densities for spatial dimensions $3-8$. The temperature dependence exhibits a crossover from sub-Arrhenius to super-Arrhenius behavior as density increases. We perform a new scaling analysis of the relaxation times to identify a density $\phi_0$ which corresponds to the ideal glass transition density for the hard sphere (or zero temperature) limit. We also estimate the (lowest) jamming density $\phi_J$, and show that for $d > 4$, $\phi_0 > \phi_J$. Comparing with theoretical calculations in \cite{mangeat2016quantitative}, we find near quantitative agreement with our estimated $\phi_0$ values (albeit with a steeper $d$ dependence for the $\phi_0$ we obtain), whereas the $\phi_J$ values we obtain are underestimated in \cite{mangeat2016quantitative}. Our results thus provide a useful benchmark for future efforts in developing quantitative theories of the glass and jamming transitions. 

\begin{acknowledgments}
\paragraph*{Acknowledgements.} 
We thank Francesco Zamponi, Patrick Charbonneau and Peter Morse for useful discussions, sharing results, and comments on the manuscript. We gratefully acknowledge Anshul Parmar for significant help in setting up the numerical investigations presented in this manuscript. 
We acknowledge the Thematic Unit of Excellence on Computational Materials Science, and the National Supercomputing Mission facility (Param Yukti) at the Jawaharlal Nehru Center for Advanced Scientific Research for computational resources. SK acknowledges support from Swarna Jayanti Fellowship grants DST/SJF/PSA-01/2018-19 and SB/SFJ/2019-20/05. SS acknowledges support through the JC Bose Fellowship  (JBR/2020/000015) SERB, DST (India).
\end{acknowledgments}

\bibliographystyle{apsrev4-1}
\bibliography{HighD}

\end{document}